\documentclass[%
 reprint,
 amsmath,amssymb,
 aps,
prb,
]{revtex4-1}

\usepackage{braket}
\usepackage{graphicx}
\usepackage{dcolumn}
\usepackage{bm}
\usepackage[normalem]{ulem} 
\usepackage{color}

\usepackage{hyperref}
\hypersetup{
    colorlinks,%
    citecolor=blue,%
    linkcolor=blue,%
    urlcolor=blue
}

\newcommand{\Ni}{(NiC$_4$S$_4)_3$}

\newcommand{\ang}[1]{#1$^{\circ}$}
\newcommand{\ntwo}{21.78$^{\circ}$}
\newcommand{\nthree}{13.17$^{\circ}$}
\newcommand{\nfour}{9.43$^{\circ}$}
\newcommand{\nfive}{7.34$^{\circ}$}
\newcommand{\nsix}{6.01$^{\circ}$}

\newcommand{\neleven}{3.14$^{\circ}$}

\newcommand{\nfithteen}{2.28$^{\circ}$}

\begin{document}

\title{Double  flat bands in kagome twisted bilayers.}

%
\author{F. Crasto de Lima, R. H. Miwa}
\affiliation{Instituto de F\'isica, Universidade Federal de Uberl\^andia, \\
        C.P. 593, 38400-902, Uberl\^andia, MG,  Brazil}%
\author{E. Su\'{a}rez Morell}
\affiliation{Departamento de F\'isica, Universidad T\'{e}cnica Federico Santa Mar\'{i}a, \\
         Valpara\'{i}so, Chile}%
\date{\today}

\begin{abstract}

We have studied how a generic bilayer kagome lattice behave upon layer rotation. We employed a Tight Binding model with one orbital per site and found (i) for low rotational angles, and at low energies, the same flat bands structure like in twisted bilayer graphene; though, for a larger value of the magic angle.
Moreover, (ii) at high energies, due to the superstructure symmetry regions, we found the characteristics three band dispersion of the kagome lattice. In the latter, its band width decreases for lower angles confining them within a few meV. Therefore, we found in twisted kagome lattice the coexistence of two sets of flat bands   in different energies and lying in different spatial regions of the bilayer system.

\end{abstract}

\maketitle

The unexpected discovery of superconductivity and correlated insulating behavior in twisted bilayer graphene (TBG)\cite{Cao2018a} at a certain "magic" angle, for which the bands structure presents very flat bands near the Fermi level\cite{Morell2010,Bistritzer2011,Li2009,TramblyDeLaissardiere2010}, has turned TBG a puzzle the community is trying to understand\cite{Tarnopolsky2018,Gonzalez2018}. The Temperature at which the material turns superconductor ($T_C$) is low, but the fact that the electrons density is thousand times smaller than in other superconducting materials, is a promising avenue to explore.  
Moreover, recently in a tetralayer structure, composed of two stacked AB bilayer graphene with a rotational angle between them,  a superconducting and correlated insulating behavior was also found by two independent groups\cite{Liu2019,Cao2019}. Additionally, in trilayer ABC structure over hexagonal boron-nitride a  Mott insulator state can be tuned by an external electric field\cite{Chen2019}. In each case the correlated behavior was related with the flat bands of the system.  Furthermore, flat bands have also been found in other non carbon based materials with interesting properties.\cite{Wu2019,Ospina2016}

Another aspect to consider is the role that interlayer coupling plays to tune the magic angle value. For instance, uniaxial pressure increases the coupling between layers thus augmenting the values of the magic angle\cite{Carr2018,Yankowitz2018}. This is not a trivial factor, it has been suggested that in this kind of system, with flat bands in their electronic structure, the critical temperature depends linearly on an attractive electron-electron interaction, and on the area of the flat band\cite{Teemu2018,Heikkila2010}. An increase in the value of the magic angle would increase the size of the Brilloiun zone and extend the area of the flat bands.

Aside from the graphene honeycomb structure thousands of other two dimensional (2D) structures have been proposed \cite{Novoselov2016,Haastrup2018}.  The kagome lattice is a network composed with the vertices and edges of the trihexagonal Archimedes tiling, it gained theoretical relevance as a realization of perfect flat bands in its electronic structure if you consider a simple nearest neighbor hopping Hamiltonian.   

\begin{figure}[ht]
\includegraphics[width=\columnwidth]{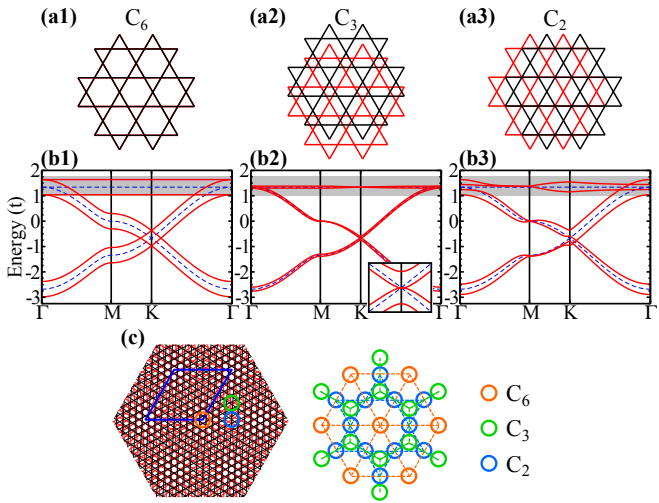}
\caption{\label{stacking-order} Different stackings of kagome bilayer (a1) $C_6$, (a2) $C_3$, (a3) $C_2$ symmetry; (b1), (b2) and (b3) the bands structure for the corresponding stacking (red lines) $C_6$, $C_3$ and $C_2$, respectively, the dashed lines are the monolayer kagome band structure; (c) identification of symmetry regions of a {\ang{3.89}} twisted kagome bilayer lattice, the blue line in the left panel shows the unit cell.}
\end{figure}

Indeed, a simple three orbitals Tight Binding (TB) model of the monolayer kagome lattice shows a flat band at high energies plus a Dirac like linear bands touching at the K points similar to graphene bands (see dashed lines in Fig.\,\ref{stacking-order}(b1)).
Several proposal of exciting new phenomena that could be found in this lattice have come out, e.g. spin liquid phases\cite{Han2012}, fractional quantum hall effect\cite{Tang2011}, quantum anomalous and spin hall effect\cite{wangPRL2013,PRB90Zhao}.

In the realm of 2D materials the kagome lattice shows up in different metal-organic frameworks (MOFs) systems. Such systems display different properties by designing combinations of metals and organic ligands \cite{ChemCommunMiguel2016}. Recently, it has been shown that Fe$_3$Se$_2$ forms a kagome bilayer structure\cite{Kang2018}, where the system is a soft ferromagnet with intrinsic anomalous Hall conductivity, and Dirac cone gaped by 30 meV as a consequence of the spin-orbit coupling. The TB fitting of the experimental results shows an  interlayer coupling of $\sim 0.3 t$, almost three times the value in TBG.

In principle in the MOFs the distance between metallic atoms within the layer might be larger than the interlayer distance  It would result in relatively larger values of the interlayer effective hopping. This property, together with a relative rotation between layers, might introduce new ingredients to the formation of flat bands in kagome twisted bilayers (KTB).

In this paper we explore the consequences that twisting a bilayer kagome lattice has to its electronic properties. We employ, based on density functional theory, a TB model with  one orbital per metallic atom and only three parameters. The calculations are done for a generic kagome lattice, nonetheless we used a $t_z/t$ ratio between inter and intra layer nearest hopping of $0.3$, similar to the value obtained in Ref [\onlinecite{Kang2018}] to fit experimental results.  Succinctly, we found TKB exhibits a larger value of the magic angle (2.28$^o$) in comparison with its counterpart TBG. The low energy electrons at this angle are spatially localized in the AA  regions, $C_6$ symmetry sites,  of the moir\'e pattern. Concomitantly, the superlattice structure composed by $C_2$ symmetry sites, which conforms a (new) $C_2$-kagome lattice, lead to a new set of kagome  bands at higher energies. Those findings show that twisted kagome lattices are characterized by two sets of  flat bands lying at  different energy and spatial regions.

The electronic structure of the bilayer kagome lattice  drastically depends on the stacking order. We study first three possible stacking orders of the bilayer structure. We label them accordingly to the symmetry of the structure, which will help us to explain below the properties of different regions of the twisted structure. In Fig.\,\ref{stacking-order}(a1)-(a3), we show the stacked AA (eclipsed) bilayer, with $C_6$ symmetry, the AB stacking with $C_3$ symmetry and another stacking obtained by shifting one of the layers half a lattice vector along the horizontal axis, this stacking has a lower $C_2$ symmetry. We calculated the electronic structure of these three structure using our TB model with distance-dependent hoppings within and between layers.

The electronic structure of a bilayer kagome lattice in the AA stacking, Fig.\,\ref{stacking-order}(b1), is similar to the structure of bilayer stacked AA graphene, we see a split of the monolayer dispersion (dashed lines) forming two identical set of bands (red lines) shifted in energy. In fact such behavior is observed in the bilayer MOF {\Ni} with AA stacking\cite{PRBdeLima2017}. The AB stacking with $C_3$ symmetry,  Fig.\,\ref{stacking-order}(b2), shows at low energy two sets of bonding-antibonding parabolic bands, similar to Bernal bilayer graphene. In the $C_2$ symmetry stacking, Fig.\,\ref{stacking-order}(b3), the degeneracy is broken at the K point with no electron-hole symmetry and linear crossings of the bands in the $K$--$\Gamma$ direction.

We can build a rotated commensurate unit cell from two stacked AA layers following the same procedure as in TBG\cite{SuarezMorell2011}, the rotation axis passes through the center of the hexagon. We obtain a superlattice unit cell where we can identify three regions with the symmetries $C_6$, $C_3$, and $C_2$. In the right side of Fig.\,\ref{stacking-order}(c) we show an schematic of how these regions are distributed over the KTB cell. Below we associate the low energy states with their spatial localization.

\begin{figure}[ht]
\includegraphics[width=\columnwidth]{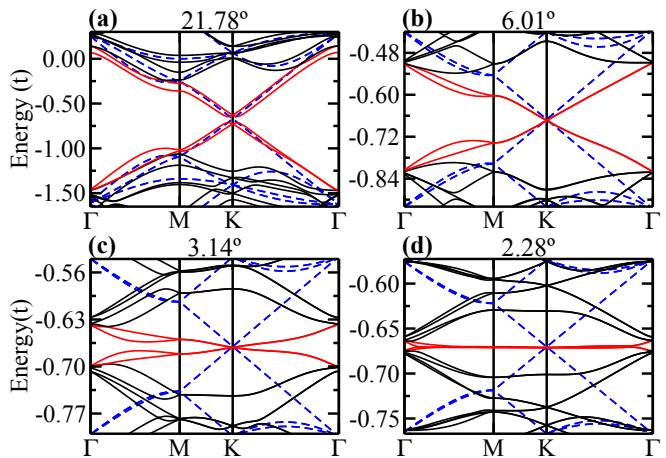}
\caption{\label{fig:kag-tw-dirac}Twisted kagome band structure close to Dirac bands for (a) {\ntwo}, (b) {\nsix}, (c) {\neleven}, (d) {\nfithteen}. The bands close to the Dirac dispersion are highlighted in red, while blue dashed lines are the monolayer dispersion.}
\end{figure}

We study now how the band structure changes when we rotate one of the kagome layers. We will concentrate first on the Dirac like bands and then on the kagome flat band.We show in Fig.\,\ref{fig:kag-tw-dirac} the band structure for four rotational angles. A gap is clearly visible in the {\ntwo} structure, for lower angles the gap disappear and the fermions velocity is renormalized until the bands are flat like in TBG. It is worth noting that the symmetry group is preserved upon twisting, and for lower twist angles the energy bands along M-K-$\Gamma$ directions are nearly degenerated, while along the  $\Gamma$-M the degeneracy is visibly lifted.

\begin{figure}[ht]
\includegraphics[width=\columnwidth]{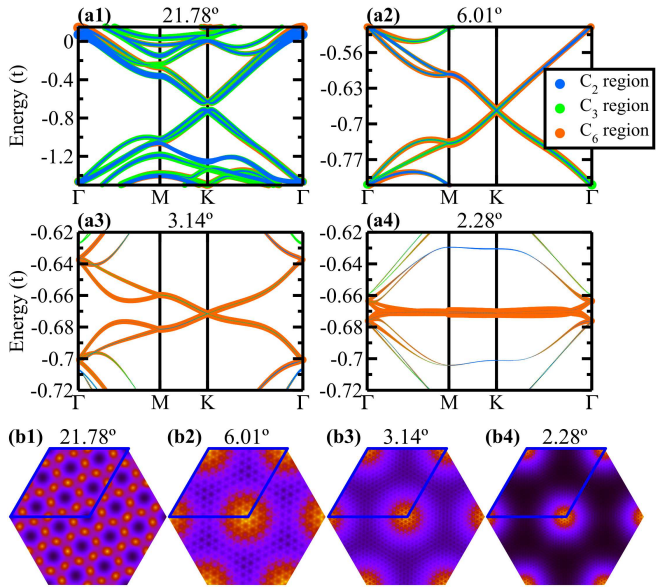}
\caption{\label{fig:kag-dirac-proj} Projected band structure in the symmetric regions for twisted angle of {\ntwo} (a1), {\nsix} (a2), {\neleven} (a3) and {\nfithteen} (a4). Real space distribution of the Dirac states for twisted angle of {\ntwo} (b1), {\nsix} (b2), {\neleven} (b3) and {\nfithteen} (b4).}
\end{figure}

    We can further understand the evolution of the Dirac bands, by looking at the projection in the band structures of the different symmetry regions Fig.\,\ref{fig:kag-dirac-proj}(a), and  the spatial localization of the Dirac states, Fig.\,\ref{fig:kag-dirac-proj}(b).  For high twisted angle, Fig.\,\ref{fig:kag-dirac-proj}(a1), the Dirac bands comprises an hybridization between the $C_6$, $C_3$ and $C_2$ regions, with most contributions coming from the $C_6$ symmetry region. With the lowering of the angle, the contribution of the $C_3$ and $C_2$ regions diminish, where for 2.28$^o$ [Fig.\,\ref{fig:kag-dirac-proj}(a4)] it has vanished. In that sense, a simplified approach tell us that as the twisting angle is lowered, and the $C_6$ regions becomes apart from each other, its hybridization diminish forming the almost flat band in the previous Dirac point, which is associated with the  presence of van Hove singularities(VHSs). In  Fig.\,\ref{fig:kag-dirac-proj}(a) we show the the angular dependence of the  VHS\cite{dos2007graphene} taking place  along the $\Gamma$-M point.     On the other hand, increasing the angle brings the $C_6$ sites closer, where its interaction with each other leads to dispersive bands. Indeed, the role the long range moir\'e potential has on the electronic properties is, currently, an area of intense research\cite{Lian2018,Zou2018,Bitan2019,Teemu2018,Fengchen2018,Mellado2018}.
    
\begin{figure}[ht]
\includegraphics[width=\columnwidth]{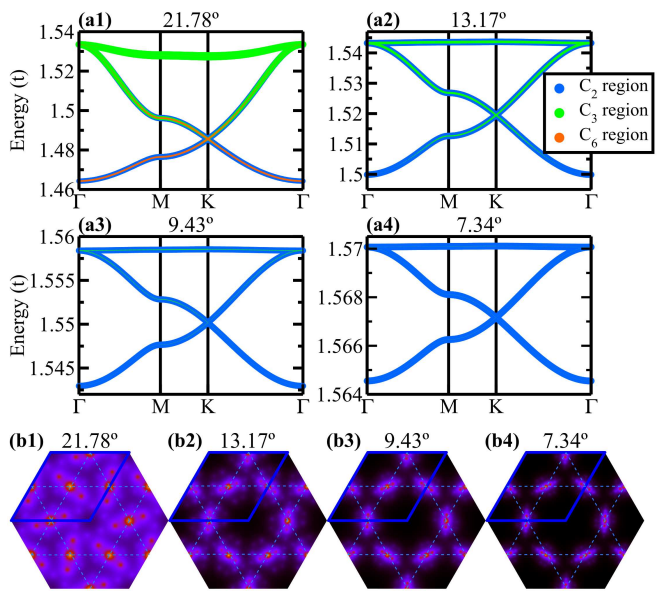}
\caption{\label{fig:kag-tw-flat} {Twisted kagome symmetric sites projected band structure close to flat bands [shaded regions in Fig.\,\ref{stacking-order}(b1)--(b3)] for (a1) {\ntwo}, (a2) {\nthree}, (a3) {\nfour} and (a4) {\nfive}}. Real space distribution of the states shown in (a), for (b1) {\ntwo}, (b2) {\nthree}, (b3) {\nfour} and (b4) {\nfive}}
\end{figure}

It is interestingly to note that, in contrast with the triangular lattice composed by the $C_6$ symmetry sites, we found  that the $C_2$ sites form a new kagome lattice embedded in the twisted system, Fig.\,\ref{stacking-order}(c). Those $C_2$ sites promote the emergence of  a new set of kagome bands above the Fermi level, namely at $\sim 1.5$\,$t$ [shaded region in Fig.\,\ref{stacking-order}(b1)]. Where the energy dispersions of those kagome bands are ruled by  the interlayer interactions of the intrinsic flat bands of each kagome monolayer. In Fig.\,\ref{fig:kag-tw-flat} we show the band structures and the orbital localization of those new kagome bands as a function of the twist angle.The evolution with the angle shows that the energy width of this three bands decreases with the angle, notice also that the high energy band become less dispersive,  strengthening the localization  on the $C_2$ sites, as the angle gets lower. Thus, similarly with what we found for the $C_6$ sites, the decrease in the band width and the flatness of these high energy bands for low angles is related with a reduction in the effective hopping between the $C_2$ regions as the distance between them increases.  

\begin{figure}[ht]
\includegraphics[width=\columnwidth]{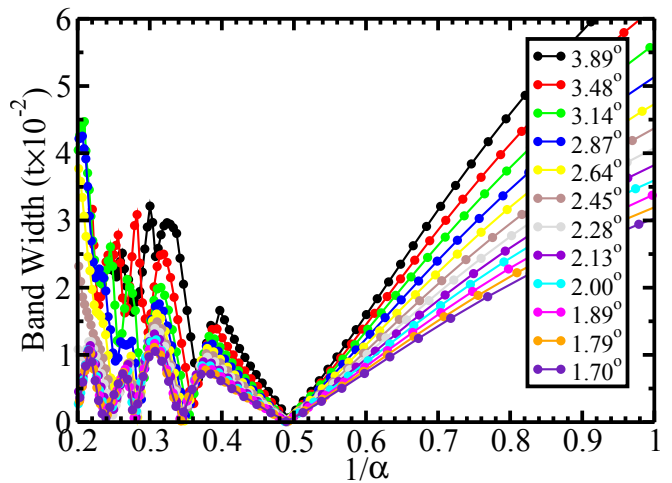}
\caption{\label{fig:band-width}Band width evolution as a function of $1/\alpha$ parameter for different angles.}
\end{figure}

Finally we shall discuss how the interlayer coupling affects the value of the magic angle.  Increasing the interlayer hopping will increase the angle for which the flat bands appear. Experimentally that can be achieved by exerting an uniaxial perpendicular pressure over the structure\cite{Yankowitz2018}. However if we normalize the interlayer hopping in terms of the intralayer hopping the sequence for magic angle will follow the same pattern. It is possible to define a dimensionless parameter\cite{Tarnopolsky2018} $\alpha=\kappa t_z/t sin(\theta/2)$, where $\kappa=2\sqrt{3}/4\pi$ for the kagome lattice, $t_z$ and t are the inter and intralayer nearest neighbor hopping and $\theta$ is the rotational angle. We plot how the bands width at the $M$ point depends on $1/\alpha$, for different angles, and obtain a sequence for the first four magic angles with basically the same pattern for all twisted structures. We are considering the width at the M point a good indicator of the flatness of low energy bands. 

With the help of the $\alpha$ parameter it is clear that to increase the value  of the magic angle we need larger ratios between the inter ($t_z$) and intra (t) layer hoppings. We have employed throughout this article a relation between $t_z/t$ of 0.3 and it gave us a magic angle of 2.28$^o$ but it is straightforward to obtain that for  $t_z/t=1$ the magic angle will be around 8$^o$.

Moreover if we assume a linear dependence of the $T_C$ on the size of the flat bands as proposed in Ref. \onlinecite{Teemu2018}, with the same coupling strength ($\lambda$) in TBG and in twisted  $Fe_3Se_2$, a simple calculation relating the size of the two Brillouin zones results in an increase of 4.7 in the value of the $T_C$ and going a little bit further, if we are able to find a material with a magic angle around 8$^o$ we will get an astonishing increase of around 50 in the critical temperature.

In conclusion we have studied a twisted bilayer kagome lattice which can be obtained in different metal-organic frameworks and found that two sets of flat bands are generated in their electronic spectrum. The origin of the low energy flat bands is the same as in twisted bilayer graphene with a characteristic triangular spatial localization, the other set of flat bands arises at the same energy of the monolayer kagome flat band and it has the typical three band dispersion of the kagome lattice, these states are localized in a different region of the superlattice unit cell  and its origin is related with an emergent kagome-like lattice in the generated moire superstructure.
We showed also that the value of the magic angle  increases in this kind of structures due to a larger interlayer/intralayer hopping ratio,  this eventually might results in an structure where the insulating correlated states appear at higher $T_C$.

\section{ACKNOWLEDGMENTS}
The authors acknowledge financial support from the Brazilian agencies CNPq, and FAPEMIG, and the CENAPAD-SP and LNCC-ScafMat for computer time. ESM acknowledges financial support from FONDECYT Regular 1170921 (Chile).

\bibliography{bib}

\end{document}